\documentclass[conference]{IEEEtran}
\IEEEoverridecommandlockouts

\usepackage{cite}
\usepackage{array}
\usepackage{amsmath,amssymb,amsfonts}
\usepackage{algorithmic}
\usepackage{graphicx}
\usepackage{textcomp}
\usepackage{xcolor}
\def\BibTeX{{\rm B\kern-.05em{\sc i\kern-.025em b}\kern-.08em
    T\kern-.1667em\lower.7ex\hbox{E}\kern-.125emX}}

\makeatletter
\newcommand{\customlabel}[2]{%
   \protected@write \@auxout{}{\string\newlabel{#1}{{#2}{\thepage}}}%
   #2%
}
\makeatother
    
\begin{document}

\title{Leveraging Agonistic-Antagonistic Coactivation in Single-Grid HDsEMG for Hand Gesture Recognition 
\thanks{

This work was supported by the NYUAD Center for Artificial Intelligence and Robotics, funded by Tamkeen under the NYUAD Research Institute Award CG010, and a grant provided by the Meem Foundation. The research was carried out on the High Performance Computing resources at NYUAD.

\textit{$^{*}$These authors are co-first authors and contributed equally to this work. At the time of this research, they were affiliated with the Engineering Division at NYU Abu Dhabi.}}
}

\author{\IEEEauthorblockN{1\textsuperscript{st} Firas Darwish$^{*}$}
\IEEEauthorblockA{\textit{Department of Statistics} \\
\textit{University of Oxford}\\
Oxford, UK \\
firas.darwish@worc.ox.ac.uk}
\and
\IEEEauthorblockN{2\textsuperscript{nd} Dhiyaa Al Jorf$^{*}$}
\IEEEauthorblockA{\textit{Department of Computer Science} \\
\textit{ETH Zurich}\\
Zurich, Switzerland \\
daljorf@ethz.ch}
\and
\IEEEauthorblockN{3\textsuperscript{rd} Costanza Armanini}
\IEEEauthorblockA{\textit{Center for Artificial Intelligence and Robotics} \\
\textit{New York University Abu Dhabi}\\
Abu Dhabi, UAE \\
ca3072@nyu.edu}
\and
\IEEEauthorblockN{4\textsuperscript{th} Eion Tyacke}
\IEEEauthorblockA{\textit{Department of Electrical and Computer Engineering} \\
\textit{New York University}\\
New York, USA \\
et1799@nyu.edu}
\and
\IEEEauthorblockN{5\textsuperscript{th} Farah E. Shamout}
\IEEEauthorblockA{\textit{Engineering Division} \\
\textit{New York University Abu Dhabi}\\
Abu Dhabi, UAE \\
fs999@nyu.edu}
}

\IEEEaftertitletext{\vspace{-1.0\baselineskip}}

\maketitle

\begin{abstract}
Surface Electromyography (sEMG) is critical for intention prediction in human-computer interfaces, such as for prosthetics control. Although deep learning models for Hand Gesture Recognition (HGR) yield excellent results, they impose high computational and hardware demands. This paper addresses this bottleneck by exploiting redundancies in agonist-antagonist muscle activity, hypothesizing that coactivations present in the sEMG signals from the extensor or flexor groups alone are sufficient for accurate HGR. We evaluate this by comparing convolutional neural networks (CNNs) trained on one muscle grid against CNN architectures trained jointly on both grids. Experiments were conducted using 16 gestures from a dataset of high-density sEMG signals from 20 subjects. The results demonstrate that the extensor grid alone achieves performance (89.5\% balanced accuracy, 0.99 AUROC) comparable to the dual-grid system (94.6\% balanced accuracy, 1.00 AUROC). Notably, even when applying slow joint fusion to capture spatial features across grids, model performance did not improve. GradCAM visualizations and anatomical analysis further indicate that the extensor region provides superior signal quality compared to the flexors.
Our findings suggest that for a base set of DoF gestures, HGR hardware requirements and computational complexity can be halved without a prohibitive loss in accuracy.
\end{abstract}


\section{Introduction}
\label{sec:introduction}

Surface Electromyography (sEMG) is a widely adopted non-invasive modality for controlling human-computer interfaces, particularly in prosthetic devices and neurorobotics \cite{prosthetic, robotic}. In these applications, Hand Gesture Recognition (HGR) models typically rely on input from either High-Density (HDsEMG) or sparse sEMG arrays. While classical Machine Learning (ML) classifiers—such as Support Vector Machines (SVMs) and Linear Discriminant Analysis (LDA) \cite{SVM, classicalML2, classicalML4} achieve high accuracy ($\geq90\%$) using time and frequency domain features, they face significant limitations as the complexity and diversity of the gesture set scale.

\begin{table*}[t]
    \centering
    \small  
    \setlength{\tabcolsep}{3pt} 
    \caption{Summary of HGR baseline accuracy performance including the approach, used architecture, number of predicted gestures $N_{\text{gest}}$, used component of the repetition signals, window length ($\text{Len}_{\text{win}}$) and overlap, sEMG type, number of parameters ($N_{\text{params}}$), training epochs ($N_{\text{epochs}}$), and number of channels ($N_{\text{ch}}$). Approaches are abbreviated as follows: Subject-Specific (SS), Pooled Dataset (PD), Test on Unseen Subjects (TUS), and Transfer Learning (TL).}
    
    \begin{tabular}{|c|c|c|c|c|c|c|c|c|c|c|c|}
        \hline
        Src. & Approach & Architecture & $N_{\text{gest}}$ & Used Portion & $\text{Len}_{\text{win}}$ & Overlap &  sEMG Type & $N_{\text{params}}$ & $N_{\text{epochs}}$ & $N_{\text{ch}}$ & Acc. \\ [0.5ex] 
        \hline\hline
        
        
        \cite{DeepHeterogenousDilationofLSTM} & SS & LSTM & 65 & First 0.5s & 300 ms & 96.7\% & HDsEMG & 540k & 28 & 128 & 82.4\% \\ [1ex] \hline
        
        \cite{HybridDeepNeuralNetworks} & SS & CNN-RNN & 17 & Entirety 
        & 100 ms & 90.0\% & Sparse & 108k & N/A & 12 & 98.0\% \\ [1ex] \hline
        
        \cite{SemgBasedHandGestureRecognitionViaDilatedConvolutionalNeuralNetworks} & SS & Dilated CNN & 17 & Entirety 
        & 100 ms & 90.0\% & Sparse & 425k & 100 & 12 & 92.5\% \\ [1ex] \hline
        
        \cite{visiontransformer1} & SS & Transformers & 65 & "Steady-State"  & 31.25 ms & 50.0\% & HDsEMG & 78k & 30 & 128 & 84.0--85.0\% \\ [1ex] \hline
        
        \cite{visiontransformer2} & SS & Transformers & 65 & Entirety & 250 ms & 87.5\% & HDsEMG & 65k & 50 & 128 & 92.0\% \\ [1ex]
        \hline\hline
        
        \cite{UnveilingEMGSemantics} & PD & CNN & 50 & Entirety 
        & 200 ms & 95.0\% & Sparse & 2.6M & 400 & 12 & 89.0\% \\ \hline
        
        \cite{atashzartowarddeepgeneralization} & PD & CNN-LSTM & 17 & Entirety
        & 300 ms & 96.7\% & Sparse & 1.1M & N/A & 12 & 79.0\% \\ [1ex] \hline
        
        \cite{UnveilingEMGSemantics} & TUS & CNN & 50 & Entirety
        & 200 ms & 95.0\% & Sparse & 2.6M & 400 & 12 & 21.1\% \\ [0.5ex]  \hline 
        
        \cite{atashzartowarddeepgeneralization} & TUS & CNN-LSTM & 17 & Entirety
        & 300 ms & 96.7\% & Sparse & 1.1M & N/A & 12 & 77.0\% \\ [1ex] \hline
        
        \cite{atashzartransienttransferlearning} & TL & d-biLSTM & 65 & First 0.5s & 200 ms & 95.0\% & HDsEMG & 78k & 200 & 128 & 73.2\% \\ [1ex]\hline
        
        \cite{eiondilatedcapsnet} & TL & dialted CapsNet & 17 & First 1s & 300 ms & 96.7\% & Sparse & 3.7M & N/A & 12 & 78.3\% \\ [1ex]\hline
    \end{tabular}
    \label{tab:baselines}
\end{table*}

To capture the complex, non-linear patterns in sEMG data, HGR research has increasingly shifted toward Deep Learning (DL) architectures. These include Convolutional Neural Networks (CNN) 
\cite{SemgBasedHandGestureRecognitionViaDilatedConvolutionalNeuralNetworks, tnsre_dl_gesture_recognition, tnsre_transfer_learning}
, Long Short-Term Memory (LSTM) networks \cite{DeepHeterogenousDilationofLSTM}, hybrid CNN-RNN architectures \cite{HybridDeepNeuralNetworks}, and vision transformers \cite{visiontransformer1, visiontransformer2}. While these methods demonstrate superior performance, they incur higher computational, data, and hardware costs. Table \ref{tab:baselines} summarizes key DL architectures used for both subject-specific and generalized HGR, serving as the baselines for this study.

Subject-specific models, trained and tested exclusively on individual data, allow for high accuracy tailored to each subject. Depending on network depth, windowing parameters, and signal type, these models achieve accuracies in the 82--98\% range. Conversely, generalized models aim for inter-subject robustness. One common strategy involves pooling data from multiple subjects for simultaneous training and testing \cite{atashzartowarddeepgeneralization}, yielding accuracies between 70\% and 90\% \cite{UnveilingEMGSemantics}. However, when tested on unseen subjects (a leave-one-subject-out approach), performance drops significantly to the 20--70\% range \cite{UnveilingEMGSemantics, atashzartowarddeepgeneralization}. To bridge this gap, transfer learning techniques have been applied. By fine-tuning a generalized model with a small amount of data from a target subject, researchers can restore accuracy to the 70--80\% range while minimizing the data collection burden \cite{atashzartransienttransferlearning, eiondilatedcapsnet}.

To address the ``black box" nature of these complex models, Explainable AI (XAI) tools such as Gradient-weighted Class Activation Mapping (GradCAM) have been adopted in sEMG analysis. GradCAM visualizes the specific input regions driving a CNN's decision-making process. For instance, \cite{atashzartowarddeepgeneralization} utilized GradCAM to analyze feature map contributions and optimize network structure. Others have leveraged it to investigate subject-specific neurophysiological biomarkers and identify key frequency ranges \cite{atashzarxmyonet}. \cite{ir.2023.28} used GradCAM heatmaps to identify high-attention regions, motivating channel reduction strategies by preserving only the most informative muscle groups.

Given the strict latency requirements for real-time HGR, mitigating the computational burden of DL is critical. A primary strategy is to reduce the model input size, lowering hardware complexity and enhancing user comfort by minimizing the sensor footprint. Previous studies have successfully optimized electrode configurations \cite{pelaez2022reducing, tomczynski2017influence, zhang2007research} or selected high-contribution channels based on post-hoc analysis \cite{ir.2023.28, atashzarxmyonet}. However, these reduction methods often require subject-specific tuning to maintain optimal performance.

\begin{figure}[ht!]
    \centering
    \includegraphics[width=0.85\linewidth]{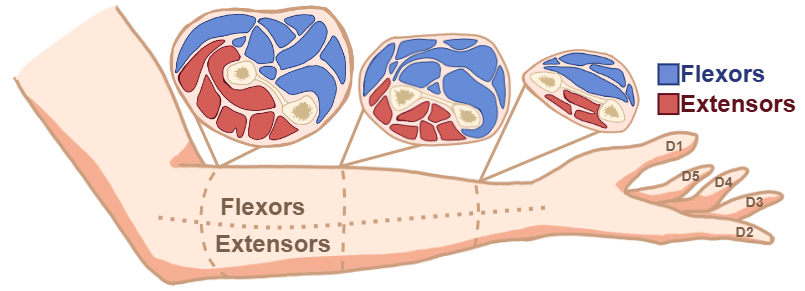}
    \caption{Front-facing forearm cross-sections at different places along the forearm with each finger labeled D1--D5}
    \label{fig:forearm}
\end{figure}

This paper proposes a novel approach to input dimensionality reduction by exploiting the inherent redundancy in agonistic-antagonistic muscle activity. Skeletal motor control is governed by antagonistic pairs, where opposing muscle groups coordinate to drive joint motion (e.g., forearm flexors driving flexion vs. extensors driving extension, as shown in Figure \ref{fig:forearm}). However, complex movements rely on muscle synergies, coordinated groupings of activations that inherently involve the coactivation of both agonists and antagonists \cite{OnTheOriginOfMuscleSynergies}. Given that these coactivation patterns are detectable via sEMG crosstalk and stabilization signals in either region \cite{Crosstalk}, we hypothesize that the signal from a single muscle group (flexor or extensor) contains sufficient discriminant information to classify a comprehensive range of gestures. To the best of our knowledge, this is the first study to investigate the systematic exclusion of an entire anatomical muscle grid based on these coactivation principles. To evaluate this hypothesis, this manuscript addresses two key questions:

\begin{enumerate} \item \textbf{\customlabel{question:1}{Extractable Redundancy}}: Can the inherent redundancy in agonistic-antagonistic muscle pairs be effectively exploited to reduce the dimensionality of the sEMG input space to one grid without compromising recognition accuracy? \item \textbf{\customlabel{question:2}{Complementary Information}}: Does the simultaneous processing of flexor and extensor signals yield complementary spatiotemporal features that improve task performance compared to when analyzing either muscle group in isolation? \end{enumerate}

\begin{figure*}[t!]
    \centering
    \includegraphics[width=0.95\linewidth]{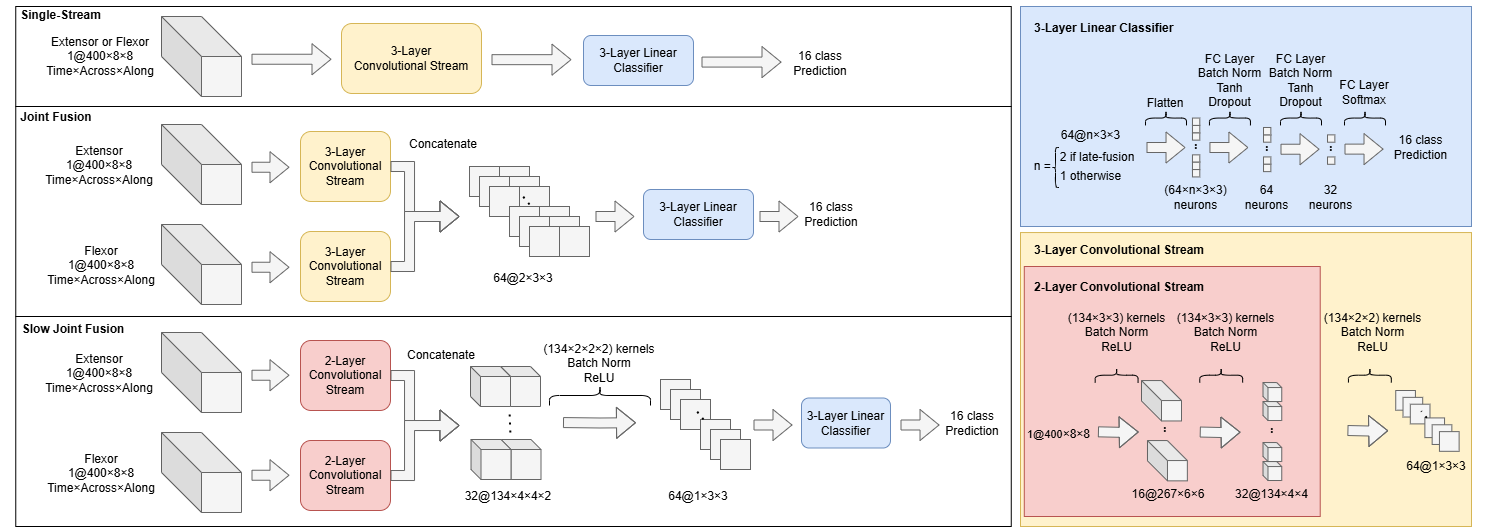}
    \caption{Model architectures (left): from top, single-stream, joint fusion, and slow joint fusion architectures. Expanded view of reused components (right): from top, 3-layer linear classifier (blue), 3-layer convolutional stream (yellow), and 2-layer convolutional stream (red).}
    \label{fig:model-structures}
\end{figure*}

To summarize, the main contributions of this work are as follows:

\begin{itemize} \item 
 We demonstrate that the extensor grid encodes redundant features that make it a sufficient input for HGR, thereby halving the computational and hardware requirements without greatly compromising accuracy.

\item 
We explain the poorer performance of the flexor-only model compared to the extensor-only model anatomically, attributing it to the muscular and adipose structure.

\item 
We employ GradCAM to interpret the decision-making process of the models. These visualizations corroborate our hypothesis by identifying regions driving accuracy.
\end{itemize}

The rest of the manuscript is organized as follows: Section \ref{sec:data}
describes the dataset and preprocessing; Section \ref{sec:methods} introduces
the methodology; Section \ref{results-discussion} presents the results and discussion and finally Section \ref{sec:conclusion} concludes the paper and presents
further avenues of research.


\section{Data}
\label{sec:data}
The HDsEMG dataset \cite{malevsevic2021database} comprises 20 subjects performing 65 gestures (5 repetitions, 5s duration). Signals were acquired at 2048 Hz using two $8\times8$ grids (10mm IED) placed on the forearm flexor and extensor muscles. We restrict our analysis to the 16 single Degree of Freedom (DoF) gestures: flexion/extension of individual fingers (D1–D5) and the wrist, plus D1 abduction/adduction and wrist rotation. This subset minimizes overfitting by simplifying decision boundaries \cite{n_classes} and enhances the interpretability of GradCAM visualizations \cite{gradcamOGpaper}, while serving as a basis set for complex motions.

We follow a modified pipeline from \cite{eionunstablecontrol} to ensure the validity of our comparative baseline analysis. Preprocessing involved hardware bandpass filtering (10–900 Hz) and notch filtering (3rd-order Butterworth) to remove power line interference. We isolate the steady-state phase by discarding the initial 1.0s (0.5s reaction time + 0.5s buffer), focusing on stable muscle contraction features rather than transient biometric identifiers \cite{atashzartransienttransferlearning}. Synchronized force data were used to validate motion direction, where positive and negative values denote extension and flexion, respectively. We implemented a 60-20-20 train-val-test split to ensure rigorous evaluation. Repetitions 1, 3, and 4 constitute the training set; repetitions 2 and 5 are split equally between validation and testing to account for fatigue effects. Signals were bandpass filtered (10–500 Hz, 4th-order), Z-score normalized (using training statistics), and rectified. Finally, data were segmented into 200ms windows with 65\% overlap, ensuring no data leakage across splits.

\section{Methods}
\label{sec:methods}

This section details the model architectures, training protocols, and the XAI approach to interpret signal redundancy.

\subsection{Model Architectures}
\label{subsec:model-struct}
To evaluate grid redundancy (Question 1) and complementary features (Question 2), we implemented three CNN architectures (Figure \ref{fig:model-structures}). All models utilize a core Convolutional Stream for feature extraction, processing a $400\times8\times8$ input (time$\times$height$\times$width) through three convolutional layers. The first two layers use $134\times3\times3$ kernels, and the third uses $134\times2\times2$. Each convolution is followed by Batch Normalization and ReLU activation, producing 64 ($1\times3\times3$) feature maps.

\begin{itemize}
\item \textbf{Single-Stream (1.77 M param):} Processes input from a single grid (extensor or flexor). The flattened 64 output feature maps are passed to a 3-layer dense classifier (576-64-32 neurons) with Tanh activation and 20\% dropout.

\item \textbf{Joint Fusion (3.55 M param):} Processes both grids via two parallel Convolutional Streams. The resulting features (64 per stream) are concatenated and fed into a wider classifier (1152-64-32 neurons). Comparing this to the Single-Stream model reveals redundancy; if performance is similar, the extra grid adds little unique information.

\item \textbf{Slow Joint Fusion (3.5 M param):} Designed to capture spatial correlations between grids. Feature maps are concatenated after the second convolutional layer, followed by a 4D convolution ($134\times2\times2\times2$) that fuses spatial information across the two streams before classification.
\end{itemize}

\noindent \textbf{Training Setup:} Models were trained on four NVIDIA RTX 8000 GPUs using the Adam optimizer ($LR=4\times10^{-5}$, cosine annealing) and cross-entropy loss (batch size 128). Training ran for 20 epochs with early stopping (patience=4). Weight decay was tuned via logarithmic grid search [$10^{-8}, 10^{-3}$].

\subsection{GradCAM Analysis}
\label{subsec:gradcam}

We apply GradCAM \cite{gradcamOGpaper} to the Joint Fusion model to quantify the contribution of each muscle grid. Heatmaps are generated from the final convolutional layer of each stream, which offers the optimal balance of spatial and semantic information. Since both streams share the loss function, their resulting gradient magnitudes are directly comparable. We generate heatmaps for all test instances and average them element-wise by gesture class, yielding 16 representative pairs (one per gesture). To assess the net influence of the extensor vs. flexor grids, we compare the mean intensity of their respective maps. Unlike standard GradCAM implementations, we omit the final ReLU step, allowing us to visualize both positive and negative contributions and providing a holistic view of how each muscle group influences the decision boundary.

\section{Results \& Discussion}\label{results-discussion}

\subsection{Extensors as Sufficient Predictors}


This section addresses Question 1 (\ref{question:1}) by evaluating the impact of excluding either the extensor or flexor grid. We analyze the observed performance differences and weigh the trade-off between the added complexity of the dual-grid system and the resulting accuracy gains.

\subsubsection{Generalized Model}\label{subsubsec:generalized-model}

We evaluate the models using the test balanced accuracy and one-vs-rest, macro-weighted Area Under Receiver Operating Characteristic Curve (AUROC). The balanced accuracy is calculated by averaging the model specificity and sensitivity to account for dataset imbalance.

Table \ref{tab:generalized-performance} shows the performance of the three generalized models after hyperparameter tuning. For the joint fusion model, the high values of the balanced accuracy (94.6\%) and AUROC (1.00) indicate balanced learning of the different gestures, and the ability of the trained model to generalize to test data, where it performs on par with CNNs with a similar number of predicted gestures ($\geq$90\%) 
\cite{HybridDeepNeuralNetworks, SemgBasedHandGestureRecognitionViaDilatedConvolutionalNeuralNetworks}. 

\begin{table}[ht!]
    \centering
    \caption{Performance of generalized models after tuning}
    \begin{tabular}{|p{15em}|c|c|}
        \hline
        \centering Model & Balanced Accuracy & AUROC \\ [0.5ex] 
        \hline\hline
        \centering Joint Fusion & 94.6\% & 1.00 \\
        \hline
        \centering Extensor Single-Stream & 89.5\% & 0.99 \\
        \hline
        \centering Flexor Single-Stream & 84.4\% & 0.98 \\ 
        \hline
        \centering Slow Joint Fusion & 94.0\% & 0.99 \\ 
        \hline
    \end{tabular}
    \label{tab:generalized-performance}
\end{table}

Similarly, the extensor single-stream models demonstrate high AUROC and balanced accuracy metrics, at almost 90\% balanced accuracy, only 5\% less than the joint fusion model. Given the much higher computational cost of the joint fusion model (i.e. double the number of trainable parameters and input size) the difference, while significant, is not as substantial for the extensor model as expected. This indicates that the extensor grid individually captures enough information to extract features associated with the flexor muscles. The flexor single-stream model’s performance at 84.4\%, while not poor, is certainly lower than that of the CNN baselines, where it is almost 10\% poorer than the joint fusion model. This substantial performance reduction warrants further investigation into the discrepancy between the flexor model and extensor model.

\begin{figure}[h]
    \centering
    \includegraphics[width=0.93\linewidth]{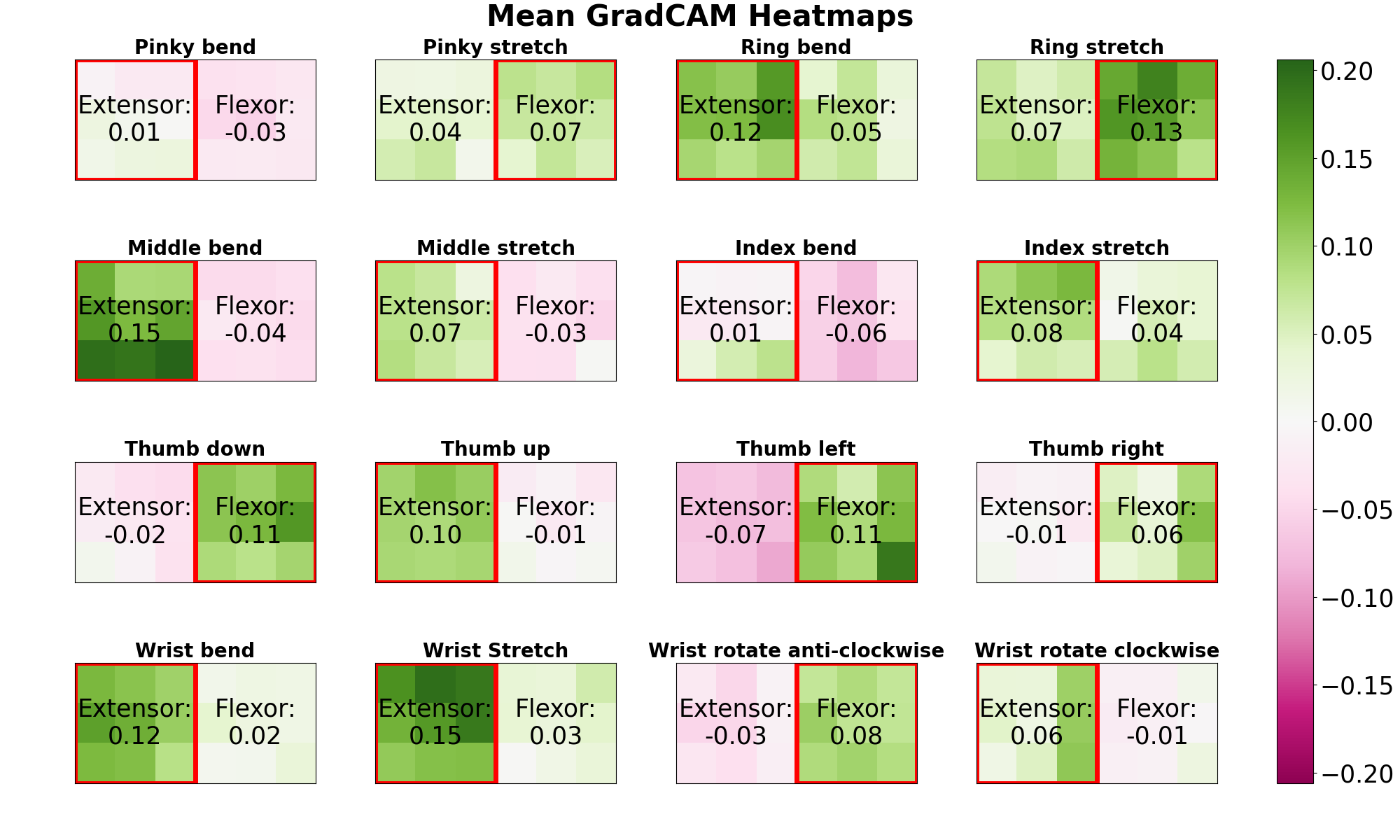}
    \caption{Average gesture-wise GradCAM heatmaps on the test dataset for the generalized joint fusion model. Red squares around the extensor (left grid) or flexor (right grid) heatmaps indicate which grid's contribution is greater for that gesture. The extensor average contribution is positive for 12/16 gestures whereas the flexor's is positive for 10/16. The extensor average contribution is greater than the flexor's 10/16 gestures.}
    \label{fig:gradcam}
\end{figure}


Figure \ref{fig:gradcam} presents the GradCAM visualizations for the joint fusion model. These results offer a method to directly compare the predictive utility of each signal type: because the fusion model optimizes a joint representation, we can interpret the magnitude of activation as the model's reliance on a specific stream. The left (extensor) and right (flexor) grids show that the average GradCAM activation for the extensor was positive for 12 gestures, whereas the flexor was positive for only 10. Furthermore, in instances where both grids were active, the extensor exhibited higher activation two-thirds of the time. This disparity suggests that the extensor signal generally encodes denser or more discernible information, causing the model to prioritize it over the flexor signal during classification.

\begin{figure}[ht!]
    \centering
    \includegraphics[width=0.93\linewidth]{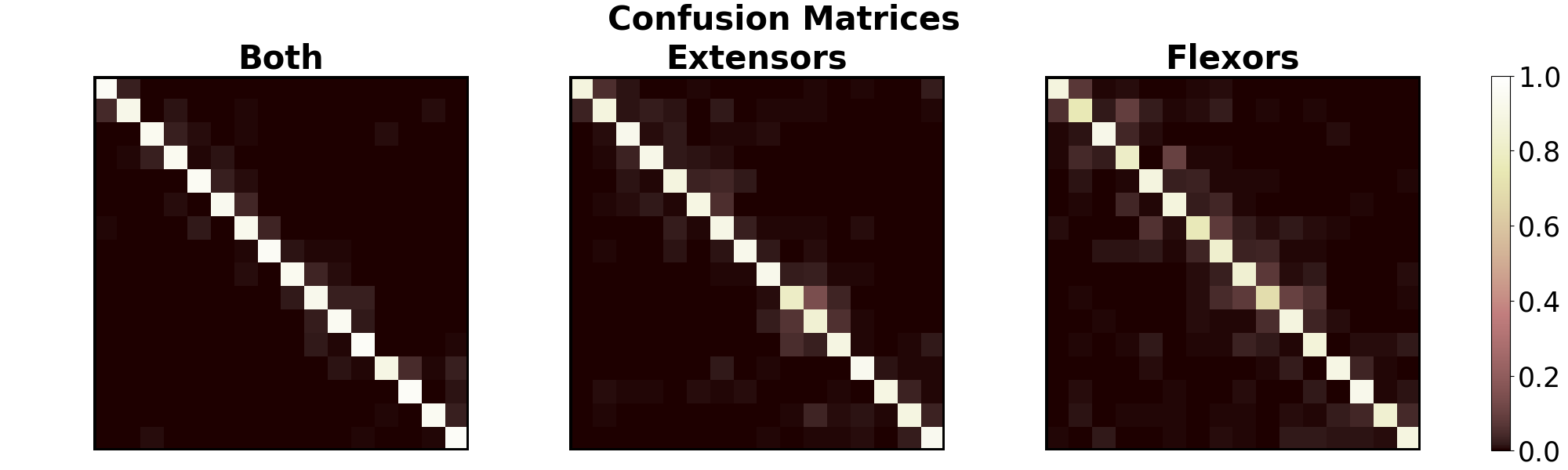}
    \caption{Confusion Matrices for the generalized models. The classes are organized from left to right and top to bottom in the order pinky bend/stretch, ring bend/stretch, middle bend/stretch, index bend/stretch, thumb up/down/left/right, wrist bend/stretch/anti-clockwise-rotate/clockwise-rotate. There is a dispersed rather than concentrated accuracy reduction for the extensor and flexor models when compared to the combined method.}
    \label{fig:confusion}
\end{figure}

Further, the confusion matrices (Figure \ref{fig:confusion}) reveal that both single-stream models are worse predictors of all gestures rather than only for their respective antagonistic motions: The flexor and extensor models achieve more dispersed precision scores in the range of 70.0--93.2\% and 80.1--94.0\% respectively for all gestures with no link between performance and gesture type when compared to the joint fusion model with a precision range of 91.8--97.1\%. To understand this disparity in performance between the single-stream models, We study aspects of the anatomy of the human arm.

Anatomical analysis of the forearm supports the earlier observations: deeper muscles contribute little to no signal to measured sEMG activity \cite{TutorialsEMGAmpliteudeEstimation}. The flexor side of the forearm is composed of many (mainly three) stacked layers of muscle shown in Figure \ref{fig:forearm}, whereas the extensor region is composed of two thinner layers. Additionally, \cite{Crosstalk} demonstrated that the magnitude of common signal due to crosstalk is much more significant among the flexor muscles ($\geq60\%$) than the extensor muscles ($\sim50\%$), which is likely attributed to the varying depths of flexor signal sources, with the deeper muscle activity having to propagate through more medium. Not to mention, while the subjects' fat composition was not recorded in the dataset for a definitive conclusion on the role adipose tissues may have played in these results, uneven forearm fat distribution may further insulate the flexor signal. Nevertheless, these factors may contribute to the poorer signal quality in the flexor sEMG data, which might explain the worse performance of its respective model as well as the more varied  performances across the different gestures.

Ultimately, these observations suggest that for the extensor grid, the answer to Question 1 (\ref{question:1}) is positive: redundancy in the form of agonistic-antagonistic coactivation is both present and detectable within the extensor grid so that the performance does not substantially degrade when the flexor grid is dropped despite the significant reduction in the input and architecture sizes. As for the flexors, the noise within the flexor signal due to the underlying anatomy detracts from the utility of redundant features and leads to performance short of the baselines when the antagonistic grid is dropped. 

To address the possibility that the observed behavior is only true for the generalized models and does not account for the effect of different subject anatomies, we further investigate whether these observations exist at a subject-specific level.

\subsubsection{Subject-Specific Models}\label{subsubsec:subject-specific-models}

To assess robustness across individual variances, we trained subject-specific models using the same hyperparameter set as the generalized models. The average performance of the subject-specific models, shown in Table \ref{tab:av-subject-specific}, closely matches that of the generalized models, where the average balanced accuracies of the subject-specific joint fusion models, extensor models, and flexor models were 95.3\%,  92.0\%, and 87.6\% respectively. 

\begin{table}[ht!]
    \centering
    \caption{Average performance \& standard deviation of subject-specific models}
    \begin{tabular}{|c|c|c|}
        \hline
        Model & Av. Balanced Accuracy & Standard Deviation \\ [0.5ex] 
        \hline\hline
        Joint Fusion & 95.3\% & 3.7\% \\
        \hline
        Extensor Single-Stream & 92.0\% & 4.7\% \\
        \hline
        Flexor Single-Stream & 87.6\% & 10.4\% \\ 
        \hline
    \end{tabular}
    \label{tab:av-subject-specific}
\end{table}

To validate the significance of these performance disparities, we evaluated the normality of the paired differences using the Shapiro-Wilk test. The normality assumption was violated for the joint fusion vs. flexor pair ($p < 0.05$), so we verified the significance using the non-parametric Wilcoxon signed-rank test, which yielded identical conclusions. For consistency across all comparisons, we report the results of the two-sided paired-samples t-tests. Effect sizes were calculated using Cohen's $d_z$. The joint fusion model demonstrated a highly significant improvement over both the extensor ($t=5.72, p < 0.001$) and flexor ($t=4.92, p < 0.001$) models, with large effect sizes (Cohen's $d_z=1.28$ and $d_z =1.10$, respectively). Furthermore, the extensor model significantly outperformed the flexor model ($t=3.18$, $p < 0.01$), exhibiting a medium-to-large effect size ($d_z =0.71$). These results confirm that the performance hierarchy observed in the generalized models holds statistically for subject-specific training.

\begin{figure}[ht!]
    \centering
    \includegraphics[width=1\linewidth]{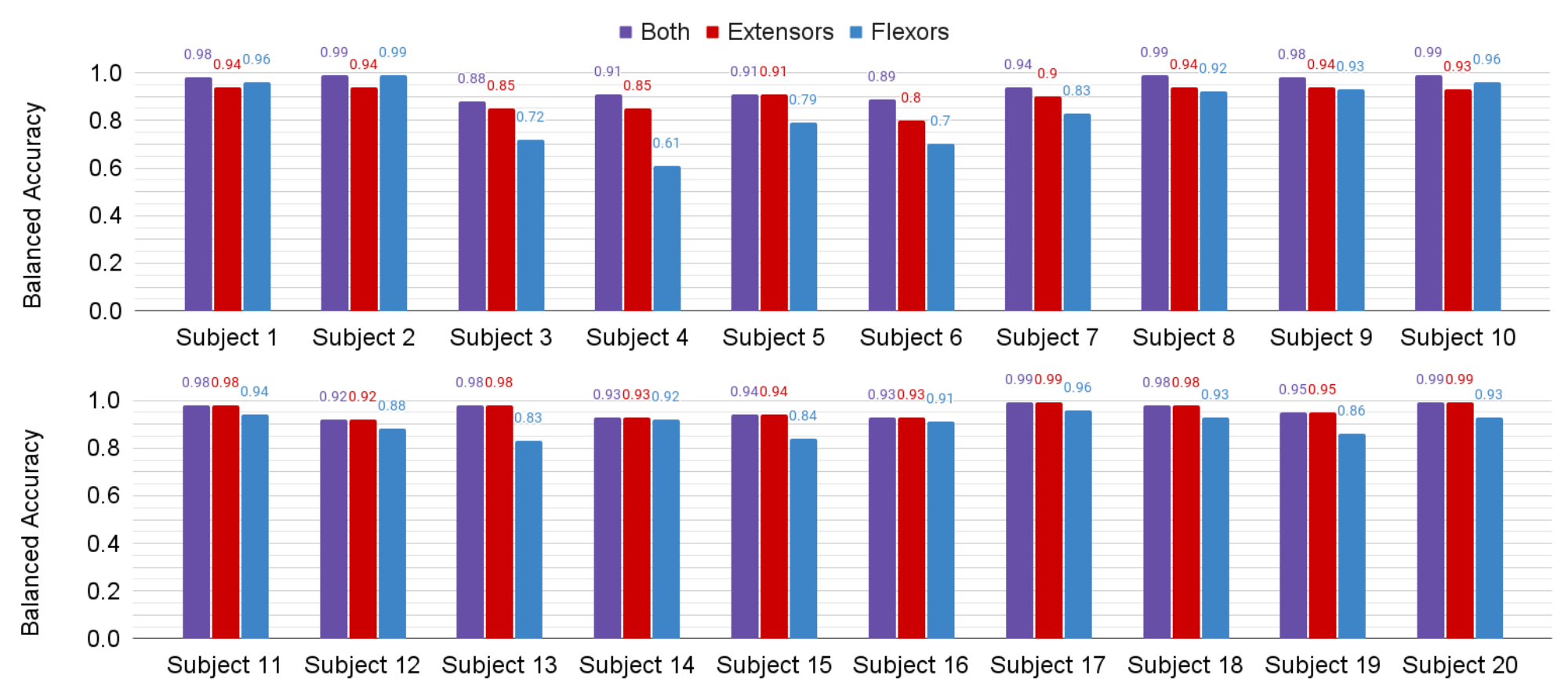}
    \caption{Subject-specific model balanced accuracies.}
    \label{fig:subject-wise}
\end{figure}

Figure \ref{fig:subject-wise} shows the balanced accuracy of the extensor and flexor single-stream models and joint fusion model on each subject. Notably, the flexor models, unlike the other two models, exhibit a lot of variance in performance across the different subjects in the range 61.0--98.5\%, further reducing the reliability of singularly using the flexor grid.

Based on the generalized models, the extensor model should perform on par with or slightly worse than the joint fusion model, whereas the flexor model shows a larger reduction in performance. Out of the 20 subjects for which subject-specific models were produced, four (1, 2, 10, and 14) deviated from the behavior outlined insofar, where the flexor grid outperformed the extensor grid by a considerable margin (2--7\%), and we interpret this by drawing from the inter-subject anatomical variations of extensor and flexor muscles. The dataset collected sEMG signals from unlabeled able-bodied men and women aged between 25 and 57 years; therefore, anatomical differences among subjects---which are not detailed in the manuscript---can significantly affect the relative performance of the extensor and flexor models \cite{SubjectIndependentsEMGPatternRecognition}. Muscle thickness \cite{RelationshipMuscleStrenghtMultiChannelSurfaceEMG}, innervation zone location \cite{TheEffectofEstimatedInnervationZoneonEMGAplitude}, skin thickness, body fat percentage, and electrode placement \cite{
eionunstablecontrol, tnsre_electrode_calibration, tnsre_continuous_learning} are known factors that cause sEMG signals to have high inter-subject variability; likewise, the unbalanced deterioration of the flexor and extensor muscles due to aging \cite{ComparisonofMaximalMuscleStrengthFlexorsExtensors} can promote intra-subject irregularity between the two grids such that the flexor grid encodes better for gesture detection. 

\begin{figure}[ht!]
    \centering
    \includegraphics[width=1\linewidth]{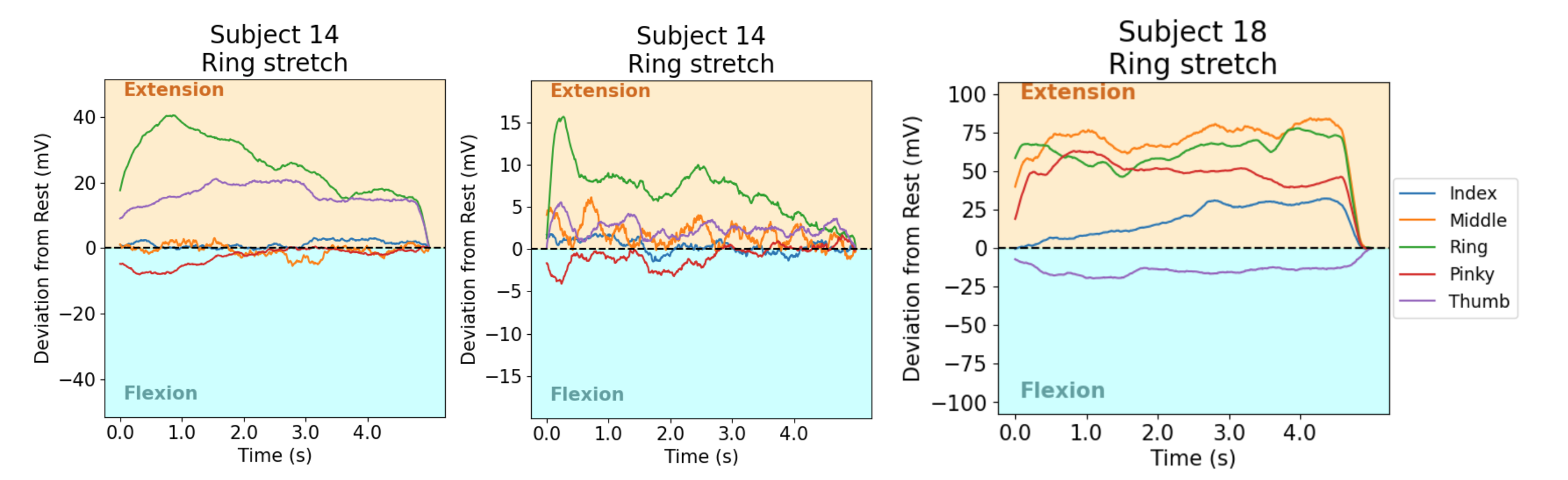}
    \caption{Force data from five of the nine available force channels to demonstrate gesture execution. Gestures are executed very differently whether by the same or different subjects. Neither force magnitudes nor stability is consistent between the three showcased repetitions. The isolated movements inadvertently involve other non-intended fingers, and even those fingers are inconsistent across the repetitions.}
    \label{fig:forces}
\end{figure}

In addition, the 9-channel forces provided by the dataset reveal differences in the gesture execution across repetitions and subjects. In Figure \ref{fig:forces}, we observe differences in the magnitudes of forces between subjects as well as how subjects maintain their gesture once in steady-state. Furthermore, these inconsistencies do not only appear when comparing inter-subject force values, where for example, subject 14 executes the same gesture with vastly different force measurements in each repetition. Repetitions for the same gesture take place sequentially with 5s of rest in between, so the observed variance in gesture execution cannot be attributed to the onset of fatigue. Finally, we discern how the one DoF motions involuntarily result in the motion of other hand components. This discrepancy is backed in \cite{ActivityPatternsofExtrinsicFingerFlexorsandExtensors}, which noted how restricting the movement to a single finger leads to muscle activity of neighboring non-instructed fingers.

Aggregating the observations across the generalized and subject specific models, we provide the following answer to Question 1 (\ref{question:1}) down to the granularity of the 16 investigated DoF gestures: For the extensor grid, the antagonistic features are extractable such that the performance of an HGR model that exclusively relies on said grid is on par with using both grids. As for the flexor grid, its anatomy results in poorer sEMG signal quality relative to the extensor muscles. This reduction negatively impacts the extractability of both agonistic and antagonistic features; therefore, the flexor signal is a poorer sole predictor in HGR tasks.

\subsection{No Improvement with Slow Joint Fusion}

Question 2 (\ref{question:2}) is concerned with whether the spatial features extractable from stacking the flexor and extensor grids and convolving across the dimension formed by that stacking add discriminative information. The joint fusion model processes the flexor and extensor grids separately through a series of convolutional layers before concatenating their formed features. Conversely, the slow joint fusion model fuses those streams earlier and convolves across the intermediary features in search of features obtainable from the spatial concatenation the two grid dimensions. Therefore, we evaluate whether there is a performance benefit to using the latter architecture.

\begin{figure}[ht!]
    \centering
    \includegraphics[width=0.70\linewidth]{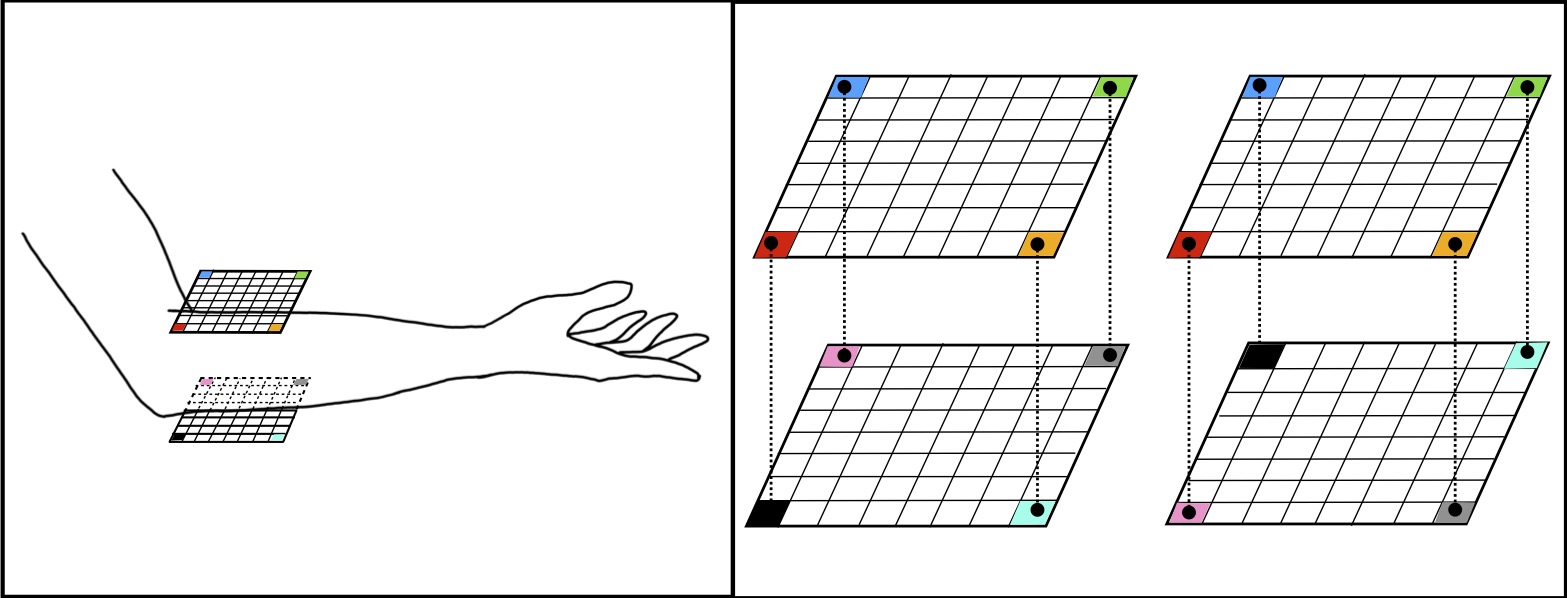}
    \caption{Reference (left): Color-coded placement of grids on forearm for reference, where top grid is flexor grid. Tested Orientations (right): rotations applied to grids before concatenation to test different spatial configurations with slow joint fusion model.}
    \label{fig:orientations}
\end{figure}

When training on data pooled from subjects 1-20, we find, as shown in Table \ref{tab:generalized-performance}, that the slow joint fusion model achieves a balanced accuracy of 94.0\%, as well as an AUROC of 0.99. These metrics are on par with the results produced by the joint fusion model. This indicates that there is no major improvement from extracting spatial features across the grids. Additionally, we tested the direct early fusion stacking of the extensor and flexor grids, as well as a mirrored stacking configuration as shown in Figure \ref{fig:orientations}. This was performed to account for any mirrored muscle activity within the forearm resulting from the relative orientations of the agonistic-antagonistic pairs, though there also was no performance improvement from these different orientations.

Based on these results, we can conclude that the spatial features unique to the stacked combination of the extensor and flexor muscle groups---regardless of grid orientation---do not provide any performance benefit. Therefore, single-stream models' performance remains comparable to fusion models. It can be concluded that fusing the two grids offers little to no incentive, given the associated multiplicative increase in complexity of HGR.

\section{Conclusion} 
\label{sec:conclusion}
This paper presents a novel approach to reducing the number of electrodes in HGR by leveraging redundant agonistic-antagonistic coactivations. We investigate two key questions: the extractability and utility of (1) redundancies between agonistic-antagonistic muscle groups and (2) beneficial features from combining the information obtained from both groups. We compare single-stream and joint fusion CNNs and demonstrate that the extensor model achieves performance comparable to that of the joint fusion model, despite halving the number of inputs and trainable parameters. Conversely, the flexor model exhibited a larger performance difference. GradCAM visualizations corroborated the observed patterns and performance differences. This performance disparity is attributed to poorer signal quality due to the anatomical structure of the forearm. These findings suggest that while both grids in an agonistic-antagonistic pair exhibit redundant coactivation, anatomical properties can affect the feature extractability from the sEMG signal. Additionally, no discriminative benefit was observed from the features extracted from combining the grids.

These findings support the hypothesis that a single grid can suffice for feature extraction. Crucially, this distinguishes informational sufficiency for classification from biological necessity for motor control; while the flexor is mechanically essential, it proved computationally redundant for this task. This approach halves the computational complexity and hardware requirements without sacrificing performance. Future work should validate this single-grid approach on transient signals, higher DoF gestures, and other agonistic-antagonistic pairs.
 
\bibliography{references}
\bibliographystyle{IEEEtran}

\end{document}